\documentclass[a4paper,superscriptaddress,showpacs,twocolumn,prl]{revtex4}%
\usepackage{amsfonts}
\usepackage{amsmath}
\usepackage{amssymb}
\usepackage{graphicx}%
\providecommand{\U}[1]{\protect\rule{.1in}{.1in}}
\begin{document}
\title{Classical and quantum correlations under decoherence}
\author{J. Maziero}
\email{jonas.maziero@ufabc.edu.br}
\affiliation{Centro de Ci\^{e}ncias Naturais e Humanas, Universidade Federal do ABC, R.
Santa Ad\'{e}lia 166, Santo Andr\'{e}, 09210-170, SP, Brazil}
\author{L. C. C\'{e}leri}
\email{lucas.celeri@ufabc.edu.br}
\affiliation{Centro de Ci\^{e}ncias Naturais e Humanas, Universidade Federal do ABC, R.
Santa Ad\'{e}lia 166, Santo Andr\'{e}, 09210-170, SP, Brazil}
\author{R. M. Serra}
\email{serra@ufabc.edu.br}
\affiliation{Centro de Ci\^{e}ncias Naturais e Humanas, Universidade Federal do ABC, R.
Santa Ad\'{e}lia 166, Santo Andr\'{e}, 09210-170, SP, Brazil}
\author{V. Vedral}
\email{vlatko.vedral@qubit.org}
\affiliation{Centre for Quantum Technologies, National University of Singapore, 3 Science
Drive 2, Singapore 117543, Singapore}
\affiliation{Department of Physics, National University of Singapore, 2 Science Drive 3,
Singapore 117542, Singapore}
\affiliation{Clarendon Laboratory, University of Oxford, Parks Road, Oxford OX1 3PU, United Kingdom}

\begin{abstract}
Recently some authors have pointed out that there exist nonclassical
correlations which are more general, and possibly more fundamental, than
entanglement. For these general quantum correlations and their classical
counterparts, under the action of decoherence, we identify three general types
of dynamics that include a peculiar sudden change in their decay rates. We
show that, under suitable conditions, the classical correlation is unaffected
by decoherence. Such dynamic behavior suggests an operational measure of both
classical and quantum correlations that can be computed without any
extremization procedure.

\end{abstract}

\maketitle

It is largely accepted that quantum mutual information is the
information-theoretic measure of the total correlation in a bipartite quantum
state. Groisman \textit{et al.} \cite{GroPoWi}, inspired by Landauer's erasure
principle \cite{Landauer}, gave an operational definition of correlations
based on the amount of noise required to destroy them. From this definition,
they proved that the total amount of correlation in any bipartite quantum
state ($\rho_{AB}$) is equal to the quantum mutual information [$\mathcal{I}%
(\rho_{A:B})=S(\rho_{A})+S(\rho_{B})-S(\rho_{AB})$, where\ $S(\rho
)=-\operatorname*{Tr}(\rho\log_{2}\rho)$ is the von Neumann entropy and
$\rho_{A(B)}=\operatorname*{Tr}_{B(A)}(\rho_{AB})$ is the reduced density
operator of the partition $A$($B$)]. Another argument in favor of the claim
that quantum mutual information is a measure of the total correlation in a
bipartite quantum state was given by Schumacher and Westmoreland
\cite{SchuWest}. They showed that, if Alice and Bob share a correlated
composite quantum system that is used as the key for a \textquotedblleft
one-time pad cryptographic system\textquotedblright, the maximum amount of
information that Alice can send securely to Bob is the quantum mutual
information of the shared correlated state.

We are interested here in the dynamics of both quantum and classical
correlations under the action of noisy environments. For these purposes, it is
reasonable to assume that the total correlation contained in a bipartite
quantum state may be separated as $\mathcal{I}(\rho_{A:B})=\mathcal{Q}%
(\rho_{AB})+\mathcal{C}(\rho_{AB})$, owing to the distinct nature of quantum
($\mathcal{Q}$) and classical ($\mathcal{C}$) correlations
\cite{GroPoWi,HenVed,Horo1,Horo2}. Some proposals for characterization and
quantification of $\mathcal{Q}$ and $\mathcal{C}$ in a composite quantum state
have appeared in the last few years \cite{GroPoWi, HenVed, OllZur, Horo1,
Winter, Piani}. The quantum correlation, $\mathcal{Q}(\rho_{AB})$, between
partitions $A$ and $B$ of a composite state can be quantified by the so-called
quantum discord, $\mathcal{D}(\rho_{AB})$,\ introduced by Ollivier and Zurek
\cite{OllZur}. Such a quantum correlation is more general than entanglement,
in the sense that separable mixed states can have a nonclassical correlation
that leads to a nonzero discord. It measures general nonclassical
correlations, including entanglement. For separable mixed states (unentangled
states) with nonzero discord, this quantum correlation provides a speed up, in
performing some tasks, over the best known classical counterpart, as was shown
theoretically \cite{Caves} and experimentally \cite{White} in a non-universal
model of quantum computation. Therefore, such a nonclassical correlation might
have a significant role in quantum information protocols. For pure states, we
have a special situation where the quantum correlation is equal to the entropy
of entanglement and also to the classical correlation. In other words,
$\mathcal{Q}(\rho_{AB})=\mathcal{C}(\rho_{AB})=\mathcal{I}(\rho_{A:B})/2$
\cite{GroPoWi, HenVed}. In this case, the total amount of quantum correlation
is captured by an entanglement measure. On the other hand, for mixed states,
the entanglement is only a part of this more general nonclassical correlation,
$\mathcal{Q}(\rho_{AB})$ \cite{OllZur, White, Caves}. A quantum composite
state may also have a classical correlation, $\mathcal{C}(\rho_{AB})$, which
for bipartite quantum states can be quantified via the measure proposed by
Henderson and one of us \cite{HenVed}. Since we assume that the total
correlation is given by the quantum mutual information and if we adopt the
definition of classical correlation given in \cite{HenVed}, $\mathcal{Q}%
(\rho_{AB})$ turns out to be identical to the definition of quantum discord in
Ref. \cite{OllZur}; in other words, $\mathcal{Q}(\rho_{AB})=\mathcal{D}%
(\rho_{AB})=\mathcal{I}(\rho_{A:B})-\mathcal{C}(\rho_{AB})$, as already noted
in Ref. \cite{Luo}.

We have identified three different kinds of dynamic behavior of $\mathcal{C}$
and $\mathcal{Q}$ under decoherence, which depend on the \textquotedblleft
geometry\textquotedblright\ of the initial composite state and on the noise
channel: $(i)$ $\mathcal{C}$ remains constant and $\mathcal{Q}$ decays
monotonically over time; $(ii)$ $\mathcal{C}$ suffers a sudden change in
behavior, decaying monotonically until a specific parametrized time, $p_{SC}$
(to be defined below), and remaining constant thereafter, while $\mathcal{Q}$
has an abrupt change in its rate of decay at $p_{SC}$, becoming greater than
$\mathcal{C}$ within certain parametrized time interval; and $(iii)$
$\mathcal{C}$\ and $\mathcal{Q}$ decay monotonically. For two-qubit states
with maximally mixed marginals we show which conditions lead to the different
types of dynamic behavior, for certain noise channels (i.e., phase flip, bit
flip, and bit-phase flip). We also recognize a symmetry among these channels
and provide a necessary condition for $\mathcal{C}$ to remain constant under
decoherence, which enables us to define an operational measure for both
classical and quantum correlations.

Let us start with the definition of classical correlation \cite{HenVed}:%
\begin{equation}
\mathcal{C}(\rho_{AB})\equiv\underset{\left\{  \Pi_{j}\right\}  }{\max}\left[
S(\rho_{A})-S_{\left\{  \Pi_{j}\right\}  }(\rho_{\left.  A\right\vert
B})\right]  , \label{CC}%
\end{equation}
where the maximum is taken over the set of projective measurements $\left\{
\Pi_{j}\right\}  $ \cite{note1} on subsystem $B$ \cite{note2}, $S_{\left\{
\Pi_{j}\right\}  }(\rho_{\left.  A\right\vert B})=%
%TCIMACRO{\tsum \nolimits_{j}}%
%BeginExpansion
{\textstyle\sum\nolimits_{j}}
%EndExpansion
q_{j}S\left(  \rho_{A}^{j}\right)  $ is the conditional entropy of subsystem
$A$, given the knowledge (measure) of the state of subsystem $B$, $\rho
_{A}^{j}=\left.  \operatorname*{Tr}_{B}\left(  \Pi_{j}\rho_{AB}\Pi_{j}\right)
\right/  q_{j}$, and $q_{j}=\operatorname*{Tr}_{AB}\left(  \rho_{AB}\Pi
_{j}\right)  $.

We consider the scenario of two qubits under local decoherence channels. The
evolved state of such a system under local environments may be described as a
completely positive trace preserving map, $\varepsilon\left(  \cdot\right)  $,
which, written in the operator-sum representation, is given by
\cite{NieChu,Dav1}
\[
\varepsilon\left(  \rho_{AB}\right)  =\sum_{i,j}\Gamma_{i}^{(A)}\Gamma
_{j}^{(B)}\rho_{AB}\Gamma_{i}^{(B)\dagger}\Gamma_{j}^{(A)\dagger}\text{,}%
\]
where $\Gamma_{i}^{(k)}$ ($k=A,B$) are the Kraus operators that describe the
noise channels $A$ and $B$.

For simplicity, let us consider a class of states with maximally mixed
marginals ($\rho_{A(B)}=\mathbf{1}_{A\left(  B\right)  }/2$), described by
\begin{equation}
\rho_{AB}=\frac{1}{4}\left(  \mathbf{1}_{AB}+\sum_{i=1}^{3}c_{i}\sigma_{i}%
^{A}\otimes\sigma_{i}^{B}\right)  ,\label{stateMMM}%
\end{equation}
where $\sigma_{i}^{k}$ is the standard Pauli operator in direction $i$ acting
on the subspace $k=A,B$, $c_{i}\in\Re$ such that $0\leq\left\vert
c_{i}\right\vert \leq1$ for $i=1,2,3$, and $\mathbf{1}_{A(B)}$ is the identity
operator in subspace $A$($B$). The state in Eq. (\ref{stateMMM}) represents a
considerable class of states including the Werner ($\left\vert c_{1}%
\right\vert =\left\vert c_{2}\right\vert =\left\vert c_{3}\right\vert =c$) and
Bell ($\left\vert c_{1}\right\vert =\left\vert c_{2}\right\vert =\left\vert
c_{3}\right\vert =1$) basis states.

\textit{Phase flip channel. }This is a quantum noise process with loss of
quantum information without loss of energy. For this channel, the Kraus
operators are given by \cite{NieChu,Dav1} $\Gamma_{0}^{(A)}=diag(\sqrt
{1-p_{A}/2},\sqrt{1-p_{A}/2})\otimes\mathbf{1}_{B}$, $\Gamma_{1}%
^{(A)}=diag(\sqrt{p_{A}/2},-\sqrt{p_{A}/2})\otimes\mathbf{1}_{B}$, $\Gamma
_{0}^{(B)}=\mathbf{1}_{A}\otimes diag(\sqrt{1-p_{B}/2},\sqrt{1-p_{B}/2})$, and
$\Gamma_{1}^{(B)}=\mathbf{1}_{A}\otimes diag(\sqrt{p_{B}/2},-\sqrt{p_{B}/2})$,
written in the subsystem basis $\left\{  |0\rangle_{k},|1\rangle_{k}\right\}
,$ $k=A,B$. We are using $p_{A(B)}$ ($0\leq p_{A(B)}\leq1$) as parametrized
time in channel $A(B)$. We consider here the symmetric situation in which the
decoherence rate is equal in both channels, so $p_{A}=p_{B}\equiv p$.

The description of the dynamical evolution of the system under the action of a
decoherence channel using the parametrized time $p$ is more general than that
using a specific functional dependence on time $t$, in the sense that it
accounts for a large range of physical scenarios. For example, for the phase
damping channel (the phase damping and phase flip channels are the same
quantum operation \cite{NieChu}), we have $p=1-\exp(-\gamma t)$, where
$\gamma$ is the phase damping rate \cite{Eberly06}.

The density operator in Eq. (\ref{stateMMM}) under the multimode noise
channel, $\varepsilon(\rho_{AB})$, has the eigenvalue spectrum:
\begin{align}
\lambda_{1} &  =\frac{1}{4}\left[  1-\alpha-\beta-\gamma\right]  ,\quad
\lambda_{2}=\frac{1}{4}\left[  1-\alpha+\beta+\gamma\right]  ,\nonumber\\
\lambda_{3} &  =\frac{1}{4}\left[  1+\alpha-\beta+\gamma\right]  ,\quad
\lambda_{4}=\frac{1}{4}\left[  1+\alpha+\beta-\gamma\right]  ,\label{lamb}%
\end{align}
with $\alpha=\left(  1-p\right)  ^{2}c_{1}$, $\beta=\left(  1-p\right)
^{2}c_{2}$, $\gamma=c_{3}$, and the von Neumann entropies of the marginal
states remain constant under phase flip for any $p$, $S\left[
\operatorname*{Tr}_{A(B)}\varepsilon\left(  \rho_{AB}\right)  \right]  =1$. To
compute the classical correlation (\ref{CC}) under phase flip, we take the
complete set of orthonormal projectors $\left\{  \Pi_{j}=\left\vert \Theta
_{j}\right\rangle \left\langle \Theta_{j}\right\vert ,j=\parallel
,\perp\right\}  $, where $\left\vert \Theta_{\parallel}\right\rangle
\equiv\cos(\theta)\left\vert 0\right\rangle +e^{i\phi}\sin(\theta)\left\vert
1\right\rangle $ and $\left\vert \Theta_{\perp}\right\rangle \equiv e^{-i\phi
}\sin(\theta)\left\vert 0\right\rangle -\cos(\theta)\left\vert 1\right\rangle
$. Then the reduced measured density operator of subsystem $A$ under phase
flip, $\widetilde{\rho}_{A}^{j}=\left.  \operatorname*{Tr}_{B}\left[  \Pi
_{j}\varepsilon(\rho_{AB})\Pi_{j}\right]  \right/  q_{j}$, will have the
following eigenvalue spectrum:
\begin{align}
\xi_{1,2}^{(j)} &  =\frac{1}{4}\left\{  2\pm\left[  2\gamma^{2}+\alpha
^{2}+\beta^{2}+\left(  2\gamma^{2}-\alpha^{2}-\beta^{2}\right)  \cos\left(
4\theta\right)  \right.  \right.  \nonumber\\
&  \left.  \left.  +2(\alpha^{2}-\beta^{2})\cos\left(  2\phi\right)  \sin
^{2}\left(  2\theta\right)  \right]  ^{1/2}\right\}  \text{,}\label{eigenval}%
\end{align}
and $q_{j}=1/2$, for $j=\parallel,\perp$. From Eq. (\ref{CC}), it follows that%
\begin{equation}
\mathcal{C}\left[  \varepsilon(\rho_{AB})\right]  =1-\underset{\theta,\phi
}{\min}\left[  S\left(  \widetilde{\rho}_{A}^{\parallel}\right)  \right]
,\label{CCM}%
\end{equation}
since $\xi_{1,2}^{(\parallel)}=\xi_{1,2}^{(\perp)}$ and hence $S\left(
\widetilde{\rho}_{A}^{\parallel}\right)  =S\left(  \widetilde{\rho}_{A}%
^{\perp}\right)  $. The classical correlation and the quantum correlation
under phase flip may be written, respectively, as%
\begin{subequations}
\begin{align}
\mathcal{C}\left[  \varepsilon\left(  \rho_{AB}\right)  \right]   &
=\sum_{k=1}^{2}\frac{1+(-1)^{k}\chi}{2}\log_{2}(1+(-1)^{k}\chi),\label{Cpf}\\
\mathcal{Q}\left[  \varepsilon\left(  \rho_{AB}\right)  \right]   &
=2+\sum_{k=1}^{4}\lambda_{k}\log_{2}\lambda_{k}-\mathcal{C}\left[
\varepsilon(\rho_{AB})\right]  ,\label{Qpf}%
\end{align}
where $\chi=\max\left(  \left\vert \alpha\right\vert ,\left\vert
\beta\right\vert ,\left\vert \gamma\right\vert \right)  $, which depends on
the relation between the coefficients $c_{i}$ in state (\ref{stateMMM}) and on
the parametrized time $p$.

$(i)$ If $\left\vert c_{3}\right\vert \geq\left\vert c_{1}\right\vert
,\left\vert c_{2}\right\vert $ in (\ref{stateMMM}), the minimum in (\ref{CCM})
is obtained by $\theta=$ $\phi=0$. The classical and the quantum correlations
under phase flip will be given in Eqs. (\ref{Cpf}) and (\ref{Qpf}),
respectively, with\ $\chi=\left\vert c_{3}\right\vert $. In this case, the
classical correlation $\mathcal{C}\left[  \varepsilon(\rho_{AB})\right]  $ is
constant (it does not depend on the parametrized time $p$) and equal to the
mutual information of the completely decohered state ($p=1$), $\mathcal{C}%
(\rho_{AB})=\mathcal{C}\left[  \varepsilon(\rho_{AB})\right]  =\mathcal{I}%
\left[  \left.  \varepsilon(\rho_{A:B})\right\vert _{p=1}\right]  $, while the
quantum correlation [Eq. (\ref{Qpf})] decays monotonically.

$(ii)$ If $\left\vert c_{1}\right\vert \geq\left\vert c_{2}\right\vert
,\left\vert c_{3}\right\vert \ $or $\left\vert c_{2}\right\vert \geq\left\vert
c_{1}\right\vert ,\left\vert c_{3}\right\vert $; and $\left\vert
c_{3}\right\vert \neq0$, we have a peculiar dynamics with a sudden change in
behavior. $\mathcal{C}$ decays monotonically until a specific parametrized
time, $p_{SC}=1-\sqrt{\left.  \left\vert c_{3}\right\vert \right/
\max(\left\vert c_{1}\right\vert ,\left\vert c_{2}\right\vert )}$, and from
then on $\mathcal{C}$ remains constant. For $p<p_{SC}$, the minimum in
(\ref{CCM}) is achieved when $\theta=\pi/4,$ $\phi=0$ (if $\left\vert
c_{1}\right\vert \geq\left\vert c_{2}\right\vert $) or $\phi=\pi/2$ (if
$\left\vert c_{1}\right\vert <\left\vert c_{2}\right\vert $), and
$\chi=\left(  1-p\right)  ^{2}\max(\left\vert c_{1}\right\vert ,\left\vert
c_{2}\right\vert )$. Thus, $\mathcal{C}$ decays monotonically. On the other
hand, for $p\geq p_{SC}$, the choice $\theta=\phi=0$ leads to the minimum in
(\ref{CCM}) and $\chi=\left\vert c_{3}\right\vert $. Then $\mathcal{C}$
suddenly becomes constant at $p=p_{SC}$, $\mathcal{C}\left[  \left.
\varepsilon(\rho_{AB})\right\vert _{p\geq p_{SC}}\right]  =\mathcal{I}\left[
\left.  \varepsilon(\rho_{A:B})\right\vert _{p=1}\right]  $, and the decay
rate of $\mathcal{Q}$ changes suddenly at $p=p_{SC}$. In Fig. 1, we depict
this peculiar behavior for a given choice of parameters and, in Fig. 2, we
show the values of the sudden change parametrized time, $p_{sc}$, as a
function of $c_{1}$ and $c_{2}$.

$(iii)$ Finally, if $\left\vert c_{3}\right\vert =0$, we have a monotonic
decay of both correlations $\mathcal{C}$ and $\mathcal{Q}$.%

%TCIMACRO{\FRAME{fhFU}{3.5198in}{2.7233in}{0pt}{\Qcb{Classical $C\left[
%\varepsilon(\rho_{AB})\right]  $ (dashed line), quantum $Q\left[
%\varepsilon(\rho_{AB})\right]  $ (solid line), and total $I\left[
%\varepsilon(\rho_{A:B})\right]  $ (dotted line) correlations under phase flip.
%We have set, in this figure, $c_{1}=0.06$, $c_{2}=0.42$, and $c_{3}=$ $0.30$ .
%For this state the sudden change occurs at $p_{sc}=0.15$, and $\QTR{cal}{Q}$
%is greater than $\QTR{cal}{C}$ for $0.09\leq p\leq0.20$. At $p=0.09$ and $p=$
%$0.20$, we have $\QTR{cal}{Q}\left[  \varepsilon(\rho_{AB})\right]
%=\QTR{cal}{C}\left[  \varepsilon\left(  \rho_{AB}\right)  \right]  =\left.
%\QTR{cal}{I}\left[  \varepsilon(\rho_{A:B})\right]  \right/  2$, as happens
%for pure states.}}{}{fig01.eps}{\special{ language "Scientific Word";
%type "GRAPHIC";  maintain-aspect-ratio TRUE;  display "USEDEF";
%valid_file "F";  width 3.5198in;  height 2.7233in;  depth 0pt;
%original-width 11.0056in;  original-height 8.4968in;  cropleft "0";
%filename 'Fig01.eps';file-properties "XNPEU";}} }%
%BeginExpansion
\begin{figure}
[h]
\begin{center}
\includegraphics[
height=2.7233in,
width=3.5198in
]%
{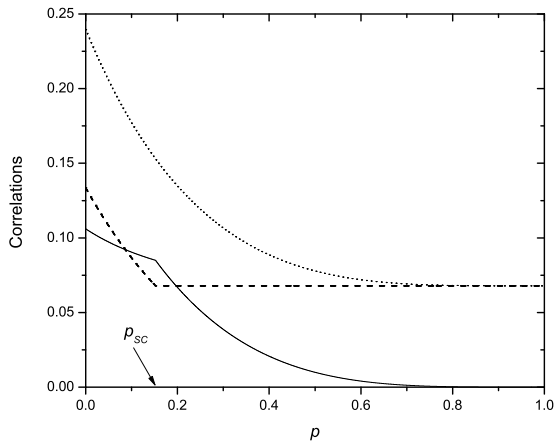}%
\caption{Classical $C\left[  \varepsilon(\rho_{AB})\right]  $ (dashed line),
quantum $Q\left[  \varepsilon(\rho_{AB})\right]  $ (solid line), and total
$I\left[  \varepsilon(\rho_{A:B})\right]  $ (dotted line) correlations under
phase flip. We have set, in this figure, $c_{1}=0.06$, $c_{2}=0.42$, and
$c_{3}=$ $0.30$ . For this state the sudden change occurs at $p_{sc}=0.15$,
and $\mathcal{Q}$ is greater than $\mathcal{C}$ for $0.09\leq p\leq0.20$. At
$p=0.09$ and $p=$ $0.20$, we have $\mathcal{Q}\left[  \varepsilon(\rho
_{AB})\right]  =\mathcal{C}\left[  \varepsilon\left(  \rho_{AB}\right)
\right]  =\left.  \mathcal{I}\left[  \varepsilon(\rho_{A:B})\right]  \right/
2$, as happens for pure states.}%
\end{center}
\end{figure}
%EndExpansion

The dynamic behavior of correlations under the phase flip channel described in
Fig. 1 is quite general. Such a sudden change in behavior occurs also when we
consider the bit flip and the bit-phase flip channels [of course under other
conditions on the $\left.  c_{k}\right.  $'s in state (\ref{stateMMM})].
Moreover, these results contradict the early conjecture that $\emph{C}%
\geq\mathcal{Q}$ for any quantum state \cite{GroPoWi,HenVed,Horo3}. Here, we
have shown that the quantum correlation may be greater than the classical one
for some states, for example $\left.  \varepsilon(\rho_{A:B})\right\vert
_{p=p_{SC}}$.

It is worth mentioning that this peculiar sudden change in behavior is a
different phenomenon from entanglement sudden death \cite{Dav1,Sudd,Eberly}.
Indeed, it seems that these correlations do not present sudden death
\cite{Werlang}.%

%TCIMACRO{\FRAME{fhFU}{2.7743in}{2.4673in}{0pt}{\Qcb{Sudden change parametrized
%time. $p_{sc}$ as a function of $c_{1}$ and $c_{2}$, for $c_{3}=0.1$, under a
%phase flip channel. In the regions where $p_{sc}=0$ or $p_{sc}=1$ there is no
%sudden change.}}{}{fig02.eps}{\special{ language "Scientific Word";
%type "GRAPHIC";  maintain-aspect-ratio TRUE;  display "USEDEF";
%valid_file "F";  width 2.7743in;  height 2.4673in;  depth 0pt;
%original-width 5.0125in;  original-height 4.4538in;  cropleft "0";
%filename 'Fig02.eps';file-properties "XNPEU";}} }%
%BeginExpansion
\begin{figure}
[h]
\begin{center}
\includegraphics[
height=2.4673in,
width=2.7743in
]%
{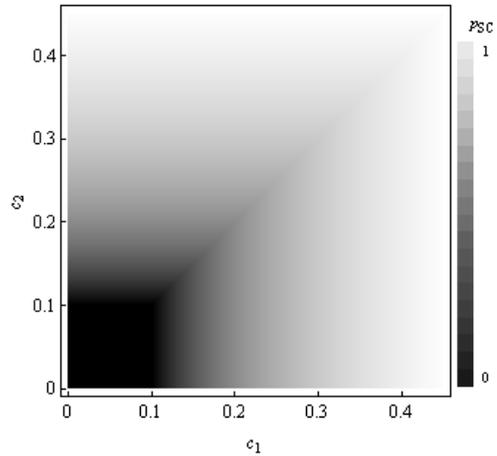}%
\caption{Sudden change parametrized time. $p_{sc}$ as a function of $c_{1}$
and $c_{2}$, for $c_{3}=0.1$, under a phase flip channel. In the regions where
$p_{sc}=0$ or $p_{sc}=1$ there is no sudden change.}%
\end{center}
\end{figure}
%EndExpansion

\textit{Bit flip channel. }The Kraus operators are \cite{NieChu,Dav1}
$\Gamma_{0}^{(A)}=diag(\sqrt{1-p/2},\sqrt{1-p/2})\otimes\mathbf{1}_{B}$,
$\Gamma_{1}^{(A)}=\sqrt{p/2}\sigma_{x}^{(A)}\otimes\mathbf{1}_{B}$,
$\Gamma_{0}^{(B)}=\mathbf{1}_{A}\otimes diag(\sqrt{1-p/2},\sqrt{1-p/2})$, and
$\Gamma_{1}^{(B)}=\mathbf{1}_{A}\otimes\sqrt{p/2}\sigma_{x}^{(B)}$. The
eigenvalue spectrum of $\varepsilon\left(  \rho_{AB}\right)  $ is given by
(\ref{lamb}), where the variables now take the form $\alpha=c_{1}$,
$\beta=\left(  1-p\right)  ^{2}c_{2}$, and $\gamma=\left(  1-p\right)
^{2}c_{3}$. The correlations can again be written as (\ref{Cpf}) and
(\ref{Qpf}). The dynamic behavior of $\mathcal{C}$ and $\mathcal{Q}$ under bit
flip is symmetrical to that for the phase flip channel (just exchanging
$c_{1}$ and $c_{3}$). Type $(i)$ dynamics is obtained when $\left\vert
c_{1}\right\vert \geq\left\vert c_{2}\right\vert ,\left\vert c_{3}\right\vert
$. Type $(ii)$ occurs for $\left\vert c_{3}\right\vert \geq\left\vert
c_{1}\right\vert ,\left\vert c_{2}\right\vert \ $or $\left\vert c_{2}%
\right\vert \geq\left\vert c_{1}\right\vert ,\left\vert c_{3}\right\vert $,
and $\left\vert c_{1}\right\vert \neq0$, with a sudden change in behavior of
$\mathcal{C}$ and $\mathcal{Q}$ at $p_{SC}=1-\sqrt{\left.  \left\vert
c_{1}\right\vert \right/  \max(\left\vert c_{2}\right\vert ,\left\vert
c_{3}\right\vert )}$. Finally, if $\left\vert c_{1}\right\vert =0$, we have
type $(iii)$ dynamics.

\textit{Bit-phase flip channel. }Now, the Kraus operators are
\cite{NieChu,Dav1} $\Gamma_{0}^{(A)}=diag(\sqrt{1-p/2},\sqrt{1-p/2}%
)\otimes\mathbf{1}_{B}$, $\Gamma_{1}^{(A)}=\sqrt{p/2}\sigma_{y}^{(A)}%
\otimes\mathbf{1}_{B}$, $\Gamma_{0}^{(B)}=\mathbf{1}_{A}\otimes diag(\sqrt
{1-p/2},\sqrt{1-p/2})$, and $\Gamma_{1}^{(B)}=\mathbf{1}_{A}\otimes\sqrt
{p/2}\sigma_{y}^{(B)}$. The variables in Eq. (\ref{lamb}) turn out to be
$\alpha=\left(  1-p\right)  ^{2}c_{1}$, $\beta=c_{2}$, and $\gamma=\left(
1-p\right)  ^{2}c_{3}$. $\mathcal{C}$ and $\mathcal{Q}$ under bit-phase flip
can again be written as (\ref{Cpf}) and (\ref{Qpf}), respectively. Once more,
the conditions for the various types of dynamics are obtained by swapping
$c_{2}$ and $c_{3}$ in the phase flip channel. For type $(ii)$ dynamics, we
now have $p_{SC}=1-\sqrt{\left.  \left\vert c_{2}\right\vert \right/
\max(\left\vert c_{1}\right\vert ,\left\vert c_{3}\right\vert )}$.

Necessary conditions for $\mathcal{C}$ to remain constant under decoherence
are the following:
\end{subequations}
\begin{equation}
\left[  \Pi_{j},\Gamma_{k}^{(B)}\right]  =0,\text{ \ }\forall\text{
\ }j,k.\label{COM}%
\end{equation}
These relations depend on the angles $\theta$ and $\phi$ that define the
minimum in (\ref{CCM}). For the channels mentioned above, $\Gamma_{0}%
^{(B)}\propto\mathbf{1}_{B}$ and $\Gamma_{1}^{(B)}\propto\sigma_{i}^{(B)}$
with $i=1$ for the bit flip, $i=2$ for the bit-phase flip, and $i=3$ for the
phase flip. Hence, condition (\ref{COM}) will be satisfied when the projective
measurements that reach the minimum in Eq. (\ref{CCM}), $\Pi_{j}$, are
performed on eigenstates of $\sigma_{i}^{(B)}$ \cite{note3}. On the other
hand, the angles $\theta$ and $\phi$ that define the minimum in Eq.
(\ref{CCM}) depend on the\ \textquotedblleft geometry\textquotedblright\ of
the initial state. When the larger component of state in Eq. (\ref{stateMMM})
is in the direction $1$, $2$, or $3$, $\mathcal{C}$ remains constant under bit
flip, bit-phase flip or phase flip, respectively.

The fact that, for a given state, the classical correlation can remain
unaffected by a suitable choice of noise channel, $\varepsilon$, immediately
suggests an operational way (without any extremization procedure) of computing
classical and quantum correlations. It could be done as follows: depending on
the state \textquotedblleft geometry\textquotedblright,\ we send its component
parts through local channels that preserve its classical correlation, so that
the quantum correlation will be given simply by the difference between the
state mutual information $\mathcal{I}(\rho_{A:B})$ and the completely
decohered mutual information, $\mathcal{I}\left[  \left.  \varepsilon
(\rho_{A:B})\right\vert _{p=1}\right]  $:%
\[
\mathcal{Q}(\rho_{AB})\equiv\mathcal{I}(\rho_{A:B})-\mathcal{I}\left[  \left.
\varepsilon(\rho_{A:B})\right\vert _{p=1}\right]  ,
\]
since $\mathcal{I}(\rho_{A:B})=\mathcal{Q}(\rho_{AB})+\mathcal{C}(\rho_{AB})$
and
\[
\mathcal{C}(\rho_{AB})=\mathcal{I}\left[  \left.  \varepsilon(\rho
_{A:B})\right\vert _{p=1}\right]  .
\]
A suitable channel for the class of states described by Eq. (\ref{stateMMM})
is chosen which satisfies condition (\ref{COM}) as discussed above.

A problem to be addressed before such a measure can be used for a general
state is to establish a protocol to find the map (if this map exists) which
leaves the classical correlation unaffected \cite{note4}. This suggests an
interesting research program to develop an operational way of investigating
the role of quantum and classical correlations in many scenarios, such as
quantum phase transitions \cite{Sarandy}, non-equilibrium thermodynamics
\cite{Vlatko}, etc.

\begin{acknowledgments}
We thank E. I. Duzzioni for discussions at the very beginning of this study.
J.M., L.C.C., and R.M.S. acknowledge the funding from UFABC, CAPES, FAPESP,
CNPq, and Brazilian National Institute of Science and Technology for Quantum
Information. V.V. acknowledges the Royal Society, the Wolfson Trust, the
Engineering and Physical Sciences Research Council (UK) and the National
Research Foundation and Ministry of Education (Singapore) for their financial support.
\end{acknowledgments}


\begin{thebibliography}{99}
\bibitem {note2}We restrict our analysis to states which have $S(\rho
_{A})=S(\rho_{B})$. In this scenario, the classical correlation computed on
measuring subsystem $A$ is equal to that given by Eq. (\ref{CC}) on measuring
subsystem $B$.
\bibitem {note3}These conditions do not depend on the representation of the
Kraus operators.
\bibitem {note4}A detailed analysis of such a measure for general states will
be presented elsewhere.
\end{thebibliography}
\end{document}